\documentclass[prl,superscriptaddress,a4paper,twocolumn]{revtex4-1}
\usepackage{geometry}
\usepackage[english]{babel}
\usepackage[latin1]{inputenc}
\usepackage{hyperref}
\usepackage{amsfonts}
\usepackage{amsmath, amssymb, amsfonts, mathrsfs}
\usepackage{graphicx}
\usepackage{xspace}
\usepackage{xcolor}
\usepackage{multirow}
\usepackage{array} 

\usepackage{float}

\newcolumntype{P}[1]{>{\centering\arraybackslash}p{#1}}
\newcolumntype{M}[1]{>{\centering\arraybackslash}m{#1}}

\usepackage[bottom]{footmisc}

\newcommand{\ket}[1]{|#1\rangle}

\begin{document}

\title{Experimental refutation of real-valued quantum mechanics under strict locality conditions}

\author{Dian Wu$^{\ast}$}
\affiliation{Hefei National Laboratory for Physical Sciences at Microscale and Department of Modern Physics, University of Science and Technology of China, Hefei, Anhui 230026, China}
\affiliation{CAS Center for Excellence and Synergetic Innovation Center in Quantum Information and Quantum Physics, University of Science and Technology of China, Shanghai 201315, China}
\affiliation{Shanghai Research Center for Quantum Sciences, Shanghai 201315, China}

\author{Yang-Fan Jiang$^{\ast}$}
\affiliation{Jinan Institute of Quantum Technology, Jinan 250101, China}

\author{Xue-Mei Gu$^{\ast}$} 
\author{Liang Huang}
\author{Bing Bai}
\author{Qi-Chao Sun}
\author{Si-Qiu Gong}
\author{Yingqiu Mao}
\author{Han-Sen Zhong}
\author{Ming-Cheng Chen}
\author{Jun Zhang}
\affiliation{Hefei National Laboratory for Physical Sciences at Microscale and Department of Modern Physics, University of Science and Technology of China, Hefei, Anhui 230026, China}
\affiliation{CAS Center for Excellence and Synergetic Innovation Center in Quantum Information and Quantum Physics, University of Science and Technology of China, Shanghai 201315, China}
\affiliation{Shanghai Research Center for Quantum Sciences, Shanghai 201315, China}

\author{Qiang Zhang}
\affiliation{Hefei National Laboratory for Physical Sciences at Microscale and Department of Modern Physics, University of Science and Technology of China, Hefei, Anhui 230026, China}
\affiliation{CAS Center for Excellence and Synergetic Innovation Center in Quantum Information and Quantum Physics, University of Science and Technology of China, Shanghai 201315, China}
\affiliation{Shanghai Research Center for Quantum Sciences, Shanghai 201315, China}
\affiliation{Jinan Institute of Quantum Technology, Jinan 250101, China}

\author{Chao-Yang Lu}
\affiliation{Hefei National Laboratory for Physical Sciences at Microscale and Department of Modern Physics, University of Science and Technology of China, Hefei, Anhui 230026, China}
\affiliation{CAS Center for Excellence and Synergetic Innovation Center in Quantum Information and Quantum Physics, University of Science and Technology of China, Shanghai 201315, China}
\affiliation{Shanghai Research Center for Quantum Sciences, Shanghai 201315, China}
 
\author{Jian-Wei Pan}
\affiliation{Hefei National Laboratory for Physical Sciences at Microscale and Department of Modern Physics, University of Science and Technology of China, Hefei, Anhui 230026, China}
\affiliation{CAS Center for Excellence and Synergetic Innovation Center in Quantum Information and Quantum Physics, University of Science and Technology of China, Shanghai 201315, China}
\affiliation{Shanghai Research Center for Quantum Sciences, Shanghai 201315, China}

\begin{abstract}
	Physicists describe nature using mathematics as the natural language, and for quantum mechanics, it prefers to use complex numbers. However, whether complex numbers are really necessary for the theory has been debated ever since its birth. Recently, it has been shown that a three-party correlation created in entanglement swapping scenarios comprising independent states and measurements cannot be reproduced using only real numbers. Previous experiments have conceptually supported the predication, yet not satisfying the independent state preparations and measurements simultaneously. Here, we implement such a test with two truly independent sources delivering entangled photons to three parties under strict locality conditions. By employing fast quantum random number generators and high-speed polarization measurements, we space-like separate all relevant events to ensure independent state preparations and measurements, and close locality loopholes simultaneously. Our results violate the real number bound of 7.66 by 5.30 standard deviations, hence rejecting the universal validity of the real-valued quantum mechanics to describe nature.   
\end{abstract}
\date{\today}
\maketitle

Quantum mechanics, in its usual formulation, is highly abstract. Its basis elements include state vectors and operators in the complex Hilbert space \cite{dirac1981principles,schrodinger1926undulatory}, and the Hilbert space of composite quantum systems is in a tensor-product structure \cite{Ismael2021}. This somehow indicates that complex numbers play a fundamental role in quantum mechanics, which is peculiar as complex numbers are simply a mathematical trick in classical mechanics. Numerous efforts have been made towards the goal of replacing complex numbers with real numbers in the Hilbert space formulation, which have lingered as evidence that the complex version of quantum theory is merely optional \cite{birkhoff1936logic,stueckelberg1959field,stueckelberg1960quantum, guenin1961quantum, dyson1962threefold, pal2008efficiency,mckague2009simulating,aleksandrova2013real}. For example, by embedding a complex density matrix $\rho$ and Hermitian observable $H$ into a double-size real Hilbert space using $\tilde{\rho}=\dfrac{1}{2}\left(\begin{array}{cc} \text{Re}(\rho) & \text{-Im}(\rho)\\ \text{Im}(\rho) & \text{Re}(\rho) \end{array} \right)$ and $\tilde{H}=\dfrac{1}{2}\left(\begin{array}{cc} \text{Re}(H) & \text{-Im}(H)\\ \text{Im}(H) & \text{Re}(H) \end{array} \right)$, the complex-valued quantum mechanics (CQM) can be simulated by real-valued quantum mechanics (RQM) through $tr(\rho H)=tr(\tilde{\rho}\tilde{H})$. This brings us an impression that RQM is physically equivalent to CQM such that quantum phenomena could be explained using only real numbers. If so, it is natural to say that complex numbers are also a convenient tool instead of an essential component in quantum mechanics.

In 1964, John Bell opened a fundamental paradigm to test the local-realist hypotheses of nature using an inequality of two-party quantum correlations \cite{bell1964einstein}. Surprisingly, Renou \textit{et al.} successfully extended this paradigm to put the real-number quantum hypotheses of nature testable for the first time using three-party quantum correlations \cite{renou2021quantum}. They observed that the three-party quantum correlations in an entanglement swapping scenario with causally independent state preparations and independent state measurements cannot be reproduced by the real hypotheses. In the real-number hypotheses, the real Hilbert space can be infinite dimensional and the Hilbert space of composite quantum systems has a tensorial structure respecting to the subsystem spaces, which further poses a requirement of independent state preparations in the experimental test \cite{renou2021quantum}. Moreover, such a tensorial structure can be expressed as a strong form of independence of space-like separated systems \cite{roos1970independence,werner1987local}. To locally observe the three-party quantum correlations, we also need to ensure the local measurements are space-like separated and thus independent. 

\begin{figure*}[!t]
	\centering
	\includegraphics[width=0.98\textwidth]{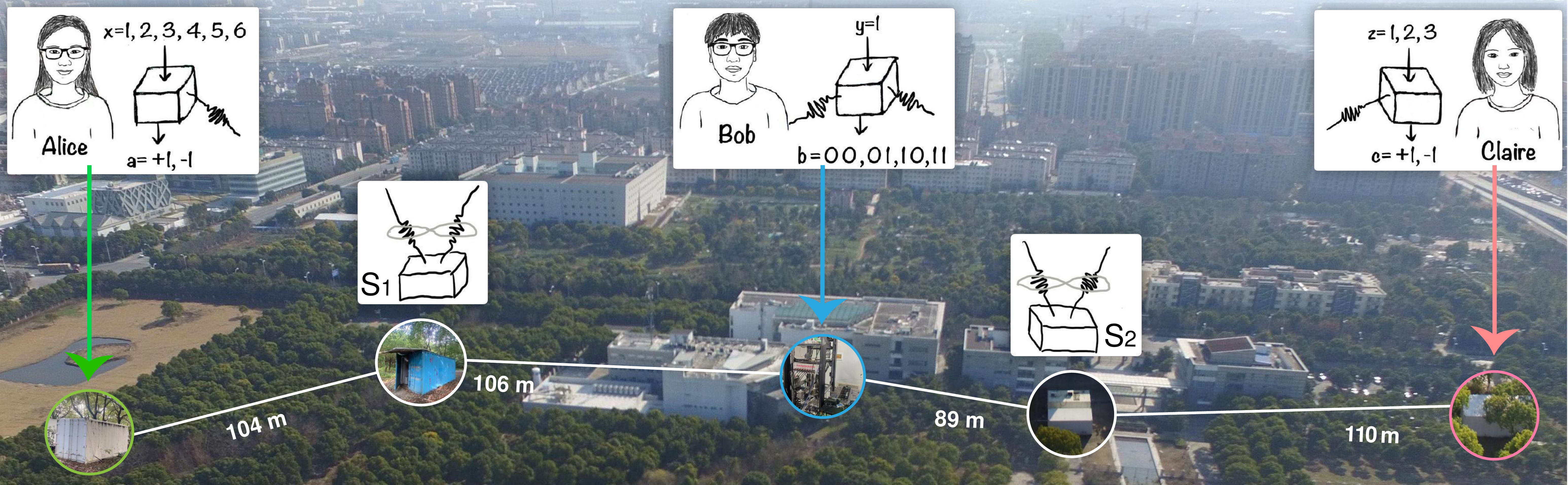}
	\caption{The non-local game to disprove RQM. Two independent sources $\text{S}_{1}$ and $\text{S}_{2}$ distribute EPR pairs between Alice and Bob, and Bob and Claire, respectively. Alice and Claire each select one of six and three inputs (labeled as $x$ and $z$) respectively and perform measurements which respectively return outcomes denoted as $a$ and $c$. Bob performs a full Bell-state measurement ($y=1$) and obtains outcome $b$ with four possibilities. The relative spatial distances between Alice-$\text{S}_1$, $\text{S}_1$-Bob, Bob-$\text{S}_2$, and $\text{S}_2$-Claire are 104 m, 106 m, 89 m, and 110 m, respectively, with optical links of distances 112.63 m, 123.84 m, 108.75 m, and 125.48 m, respectively. 
	} \label{Fig:concepts}
\end{figure*} 
Similar to the history of Bell tests, where progressively more sophisticated experiments were performed to refute local-realist hypotheses by closing the potential loopholes \cite{weihs1998violation,scheidl2010violation,rowe2001experimental,larsson2014bell,giustina2013bell,giustina2015significant,hensen2015loophole,shalm2015strong,rosenfeld2017event,li2018test}, we note that the recent experimental tests on the real-number hypotheses have not yet satisfied the requirement of independent quantum preparations and measurements simultaneously. Based on a superconducting chip \cite{chen2021ruling}, all qubits were placed in a linear array and have direct nearest-neighbour couplings. The state preparation and measurements were not strictly space-like separated, albeit it has used a I-shape transmon qubit design \cite{barends2013coherent} to increase the spacing. Based on an photonic system \cite{li2021testing}, two different lasers and proper wave-plates on the same optical table were used for independent state preparations and local measurements, respectively. No space-liked separation was enforced in that experiment. For a foundational rejection of RQM, the causality independence is crucial \cite{renou2021quantum}. Moreover, potential loopholes may be exploited by RQM to replicate the prediction of CQM. Here and for the rest of this paper, RQM refers to the real-number hypotheses of quantum mechanics.

In this work, we intend to refute RQM under strict locality conditions in order to satisfy the independent quantum preparations and measurements, and close the locality loophole simultaneously. By employing two truly independent photon sources, fast quantum random number generators \cite{qi2010high,nie2015generation} (QRNGs), high-speed single-photon polarization measurements, and space-like separation all relevant events, we demonstrate that RQM cannot reproduce the predication of CQM, thus disproving the universal validity of RQM to describe nature.

The Bell-like test proposed by Renou \textit{et al.} is depicted in Fig. \ref{Fig:concepts}, involving two independent sources and three separated players, Alice, Bob, and Claire. There, the two independent sources prepare two Einstein-Podolsky-Rosen (EPR) entangled states \cite{einstein1935can} such that one distributes to Alice and Bob, and the other to Bob and Claire. Then Alice and Claire independently perform their measurement $A_{x}$ and $C_{z}$ with random inputs $x\in\{1,2,3,4,5,6\}$ and $z\in\{1,2,3\}$, and produce outcomes as $a,c=\pm1$. Bob performs a single measurement labeled as input $y$ that returns four outcomes $b=b_{1}b_{2}\in\{0,1\}^2$. The three-party correlation is defined as the weighted sum of a input-output probability distribution $p(abc|xz)$, given as \cite{renou2021quantum}\begin{equation}\mathcal{F}=\sum_{abc,xz}^{}\omega_{abc,xz} p(abc|xz),
\end{equation}where $w_{abc,xz}=\pm1$ are the weights, $abc\in\{0,1\}^{\otimes^{4}}$ are the bit strings of the measurement results, and $xz\in\{11,12,21,22,13,14,33,34,52,53,62,63\}$  are the 12 combinations of the settings $x$ and $z$.

Given that the two EPR pairs in state $\ket{\Phi^{+}}=(\ket{00}+\ket{11})/\sqrt{2}$, Alice's and Claire's measurements are set as $A_{x}\in\{Z\pm X, Z\pm Y, X\pm Y\}/\sqrt{2}$ and $C_{z}\in\{Z,X,Y\}$, respectively. Here $Z$, $X$, and $Y$ are the standard Pauli operators. Bob conducts a full Bell state measurement (BSM), which will give four outcomes equivalently to the four standard Bell states $\ket{\Phi^+}$, $\ket{\Psi^+}$, $\ket{\Phi^-}$, and $\ket{\Psi^-}$. For CQM, the three-party correlation can reach a maximal value of $\mathcal{F}=6\sqrt{2} (\approx8.49)$. However, for RQM, $\mathcal{F}$ is upper bounded by 7.66, which sits between the classical limit of 6 and the quantum limit of $6\sqrt{2}$. This paves a way for experimentally distinguishing CQM from RQM.
\begin{figure*}[!t]
	\centering
	\includegraphics[width=0.95\textwidth]{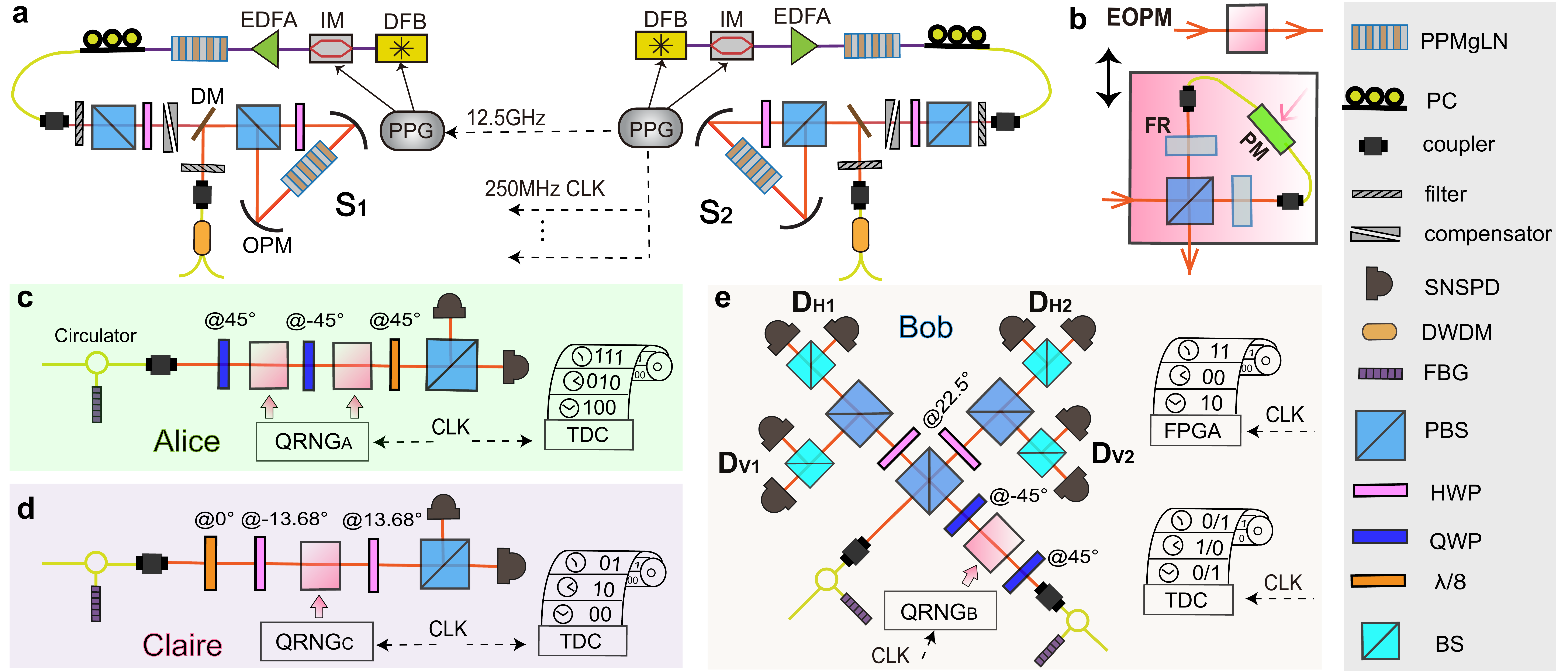}
	\caption{Experimental setup. \textbf{a}, In each source, a 1558 nm distributed feedback (DFB) laser is frequency-doubled in a periodically poled MgO doped lithium niobate (PPMgLN) crystal after passing through an erbium-doped fiber amplifier (EDFA), in order to produce 779 nm laser pulses. The second PPMgLN crystal in a polarization-based Sagnac loop is then pumped to create polarization-entangled photon pairs. In $\text{S}_2$, the pulse pattern generator (PPG) offers 250 MHz synchronization signals (CLK, black dashed lines) to all parties, and 12.5 GHz signals to trigger the PPG at $\text{S}_1$ (more details see Supplementary). \textbf{b}, An electro-optic polarization modulation (EOPM) consisting of a polarization beam splitter (PBS), two Faraday rotators (FR), and a fiber-coupled electro-optic phase modulator (PM) is implemented for modulating photon's polarization (see Supplementary). \textbf{c}, Alice measures her photon polarization in one of six bases ($A_{x}\in\{Z\pm X, Z\pm Y, X\pm Y\}/\sqrt{2}$), determined by the co-located quantum random number generator ($\text{QRNG}_{A}$) and a fast polarization modulator (which consists two EOPMs and fixed wave-plates), and records measurement outcomes using a time-to-digital converter (TDC). \textbf{d}, Similar to \textbf{c},  Claire performs her measurement setting in one of three bases ($C_{z}\in\{Z,X,Y\}$), and records the outcomes in a synchronized TDC. \textbf{e}, A partial Bell-state measurement (BSM) analyzer is constructed by three PBSs, two HWPs, four 50:50 beam splitters (BSs), and eight superconducting nanowire single photon detectors (SNSPDs). In front of the first PBS, a polarization modulator (containing EOPM and QWPs) situated there. Bob then performs the full BSM of four possible outcomes which are randomly decided by the outcomes of $\text{QRNG}_{B}$ and random photon clicks at SNSPDs. The choice of $\text{QRNG}_{B}$ is recorded by a TDC, and the photon detection results are analyzed in real time and recorded by a field-programmable gate array (FPGA). IM, intensity modulator; PC, polarization controller; DWDM, dense wavelength-division multiplexer; DM, dichroic mirrors; OPM, off-axis parabolic mirrors; FBG, fiber Bragg grating; QWP, quarter-wave plate; $\lambda/8$, eighth wave-plate.} \label{Fig:setup}
\end{figure*}   

Our detailed experimental setup, located in Shanghai campus of the University of Science and Technology of China, is illustrated in Fig. \ref{Fig:setup}. In our experiment, all events are synchronized by a pulse pattern generator (PPG) in source $\text{S}_{2}$, which provides a 12.5 GHz signal for the PPG in source $\text{S}_{1}$, and a 250 MHz signal as a time reference to synchronize all operations and measurements at the three players' locations (see Supplementary and Ref. \cite{sun2019experimental} for more details). Driven by the 250 MHz signal, two distributed feedback (DFB) lasers each emit a 2 ns laser pulse (central wavelength 1558 nm) which is further shorten to 80 ps by an intensity modulator (IM). Then, the pulse is fed into a PPMgLN crystal for second harmonic generation to create a 779 nm pump laser. With such a pump laser pumping the PPMgLN crystal in a polarization-based Sagnac loop, two EPR pairs in the form of $\ket{\Phi^{+}}=(\ket{HH}+\ket{VV})/\sqrt{2}$ are created by the Type-0 spontaneous parametric down conversion (SPDC) process, where $\ket{H}$ and $\ket{V}$ are the horizontal and vertical polarization, respectively, shown in Fig. \ref{Fig:setup}a. Importantly, the 250 MHz laser pulses can raise its electric current from much below to well above the lasing threshold such that the phase of each generated laser pulse is randomized in each source \cite{sun2019experimental}. The two SPDC processes in the two space-like separated sources are therefore uncorrelated, which strongly guarantees the independent state preparations to falsify RQM \cite{renou2021quantum}. 
\begin{figure*}[!t]
	\centering
	\includegraphics[width=1\textwidth]{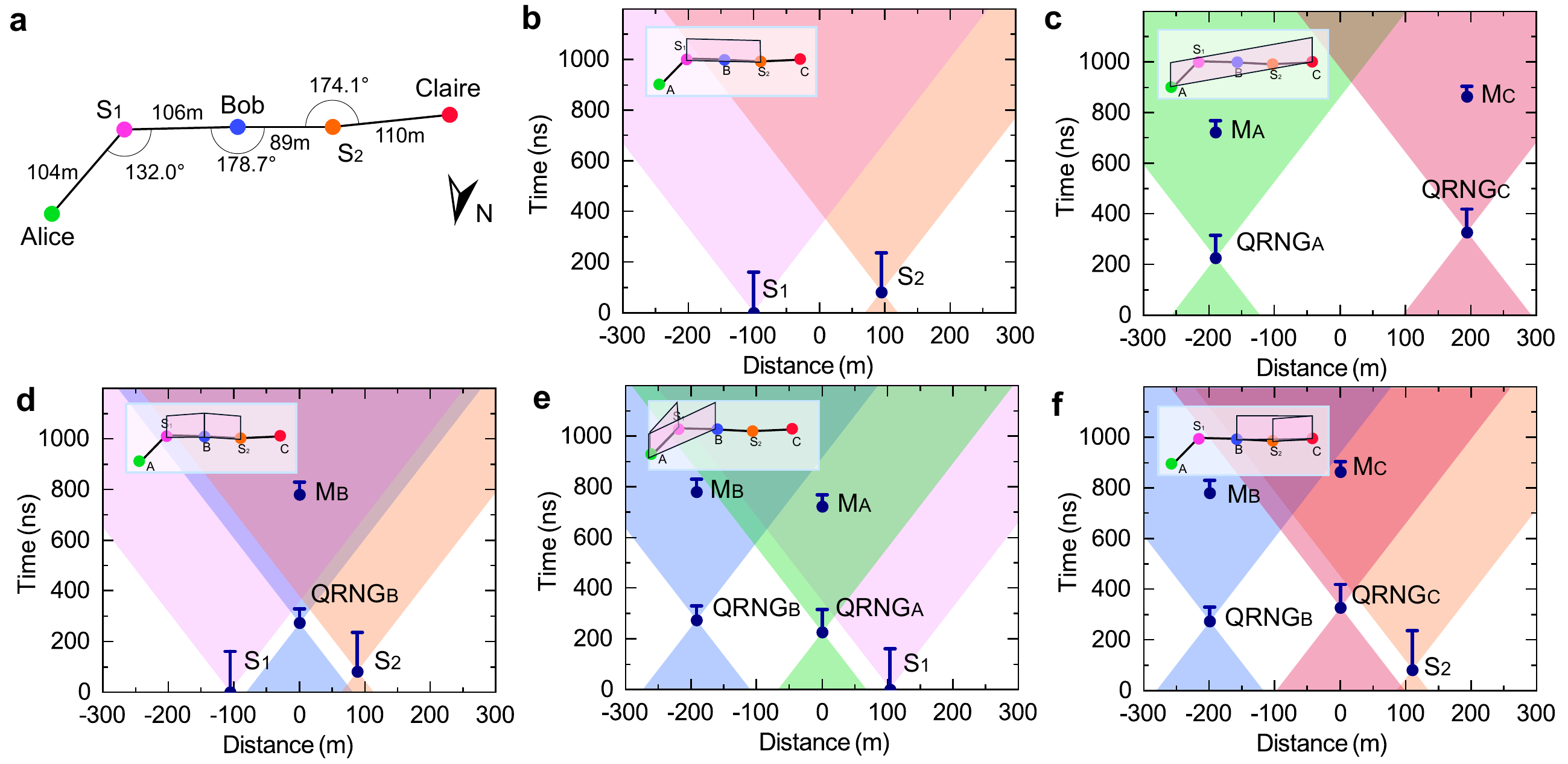}
	\caption{Space-time configuration of relevant events in our experiment. Origins of the axes are displaced to reflect the relative space and time difference between them. \textbf{a}, Relative spatial configuration of the two independent sources ($\text{S}_{1}$ and $\text{S}_{2}$) and three players (Alice, Bob, and Claire). \textbf{b}, Space-like separation between state emission events from sources $\text{S}_{1}$ and $\text{S}_{2}$. \textbf{c}, Space-like separation between setting choice events $\text{QRNG}_{A}$ and $\text{QRNG}_{C}$, and space-like separation between setting choice event $\text{QRNG}_{A}$($\text{QRNG}_{C}$) and measurement event $\text{M}_{C}$($\text{M}_{A}$). \textbf{d}, Space-like separation between setting choice event $\text{QRNG}_{B}$ and state emission events from sources $\text{S}_{1}$ and $\text{S}_{2}$ . \textbf{e}, Space-like separation between setting choice events $\text{QRNG}_{A}$ and $\text{QRNG}_{B}$, and space-like separation between setting choice event $\text{QRNG}_{A}$($\text{QRNG}_{B}$) and measurement event $\text{M}_{B}$($\text{M}_{A}$) shown on the left-side of the vertical axis, with the state emission event $\text{S}_{1}$ on the right-side. \textbf{f}, Similar to \textbf{e}, $\text{QRNG}_{C}$ is the setting choice event of Claire and the state emission event is $\text{S}_{2}$. Blue vertical bars indicate the time elapsing for events, with start and end marked by circles and horizontal line, respectively. All the time-space relations are drawn to scale. Therefore, further time-space relations can be inferred; for example, the space-like separation of $\text{QRNG}_{A}$-$\text{S}_{2}$ and $\text{QRNG}_{C}$-$\text{S}_{1}$ is implied by \textbf{e}, \textbf{f}. For details see Supplementary.} \label{Fig:spacetime}
\end{figure*}

To vary the direction of local polarization analysis for Alice and Claire, high-speed polarization analyzers are implemented with fixed waveplates, electro-optic polarization modulators (EOPMs, shown in Fig. \ref{Fig:setup}b), polarization beam splitter (PBS), and two superconducting nanowire single-photon detectors (SNSPDs), which are shown in Fig. \ref{Fig:setup}c and d (for specific configurations refer to Supplementary). The setting choices for $A_{x}$ and $C_{z}$ are determined by the fast QRNGs situated at Alice's ($\text{QRNG}_{A}$) and Claire's ($\text{QRNG}_{C}$) stations in real time, respectively. Note that although $\text{QRNG}_{A}$ outputs eight random bits and $\text{QRNG}_{C}$ gives four random bits, we can discard unnecessary ones to have six and three setting choices. For Bob, with standard linear optical Bell-state analyzer, only two of four Bell states can be distinguished \cite{weinfurter1994experimental,calsamiglia2001maximum}, e.g., a partial BSM for $\ket{\Phi^{\pm}}$ with a successful rate of 50\%. One could also have a partial BSM for $\ket{\Psi^{\pm}}$. That means we can combine two partial BSMs as a full BSM with an effective efficiency of 50\%. The full BSM is implemented in two steps: (1) \textit{pre-measure}, i.e., the entangling measurement projects the quantum states to a two-dimensional subspace determined by a fast QRNG; (2) \textit{measure}, i.e., the subspace is further projected to one unique Bell state in the subspace registered by random photon clicks. With the arrangement in Fig. \ref{Fig:setup}e, we describe how the two steps work. In \textit{pre-measure}, a QRNG at Bob's station ($\text{QRNG}_{B}$) outputs a random bit to trigger the EOPM sitting between two quarter-wave plates (QWPs). This will project the incoming quantum state onto a two-dimensional subspace in either the partial BSM of $\ket{\Phi^{\pm}}$ or the partial BSM of $\ket{\Psi^{\pm}}$, depending on $\text{QRNG}_{B}$. In \textit{measure}, four pseudo-photon-number-resolving detectors (PRNRDs, denoted as $\text{D}_{\text{H1}}$, $\text{D}_{\text{V1}}$, $\text{D}_{\text{H2}}$ and $\text{D}_{\text{V2}}$) constructed by a 50:50 beam splitter and two SNSPDs, are used to register random photon clicks as well as mitigate the multi-photon effect in the experiment. Coincidence detection between either $\text{D}_{\text{H1}}$ and $\text{D}_{\text{H2}}$ or $\text{D}_{\text{V1}}$ and $\text{D}_{\text{V2}}$ indicates that the pre-measure subspace collapses to a unique Bell state in either $\ket{\Phi^{+}}$ or $\ket{\Psi^{+}}$, while coincidence detection between either $\text{D}_{\text{H1}}$ and $\text{D}_{\text{V2}}$ or $\text{D}_{\text{V2}}$ and $\text{D}_{\text{H2}}$ shows that the subspace collapses to a Bell state in either $\ket{\Phi^{-}}$ or $\ket{\Psi^{-}}$. Hence, we have the full BSM that gives four measurement outcomes. The single photon detection events are analyzed in real time and recorded by a field-programmable gate array (FPGA). In addition, each implemented setting was recorded locally at each station using a time-to-digital converter (TDC) and all locally stored data are collected by a separate computer that evaluates the three-party correlation.

To satisfy the requirements of independent state measurement and locality constraint, it is essential to space-like separate each local setting choice (labeled as $\text{QRNG}_{A}$, $\text{QRNG}_{B}$, and $\text{QRNG}_{C}$) from the measurement of other stations (denoted as $\text{M}_{A}$, $\text{M}_{B}$, and $\text{M}_{C}$), as well as from the photon emission (labeled as  $\text{S}_{1}$ and $\text{S}_{2}$). By separating the setting choice events space-like from the events of entanglement creation in the sources, we ensure independent measurement. We close the locality loophole similarly by separating setting choice events on one side space-like from the measurement events on other sides. Keep in mind that it is impossible to rule out all loopholes \cite{brunner2014bell}, as the hidden variables could have been correlated at the birth of the universe. However, it is reasonable to assume that they are created together with the creation of EPR states as in the so-called loophole-free Bell tests \cite{giustina2015significant,hensen2015loophole,shalm2015strong,rosenfeld2017event}. We therefore consider the photon emission in the source to be the origin of a hidden variable in our experiment. We precisely characterize all relevant events (detailed analysis in the Supplementary) and describe them in Fig. \ref{Fig:spacetime}, which ensures the truly independent quantum preparation and measurement, and close locality loopholes simultaneously.

\begin{figure}[!t]
	\centering
	\includegraphics[width=0.47\textwidth]{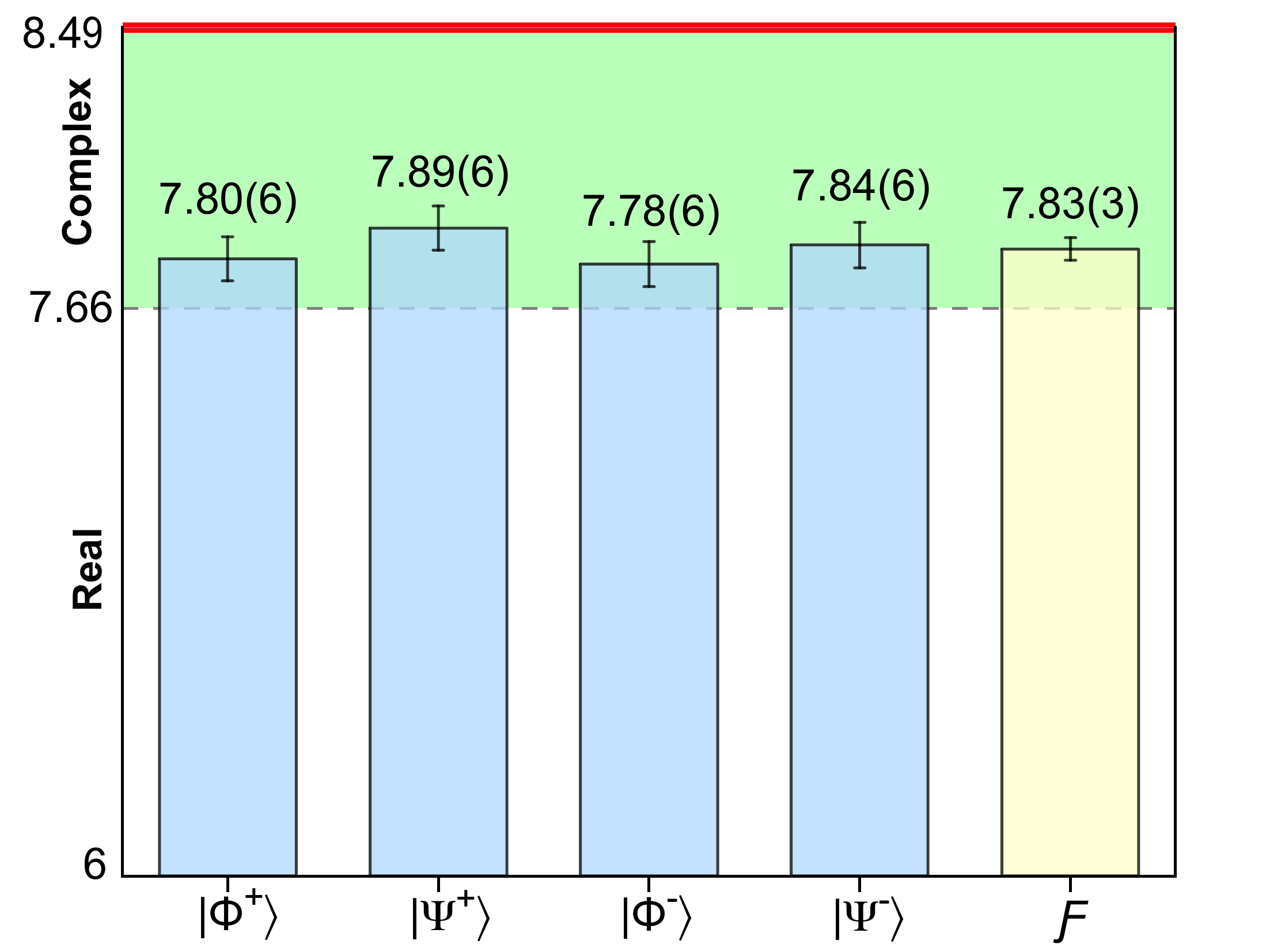}
	\caption{Experimental results. The upper bounds of the three-party correlation for classical mechanics, RQM, and CQM are 6, 7.66, and 8.49, respectively. The results for Bob's four measurement outcomes are 7.80(6), 7.89(6), 7.78(6), and 7.84(6). The average result is 7.83(3), which violates the real bound of 7.66 by 5.30 standard deviations. Error bars indicate one standard deviation and are calculated from the experimentally detected measurement events with propagated Poissonian counting statistics.} \label{Fig:results}
\end{figure}

To verify the experimental results, we perform tomographic measurements for the two independent EPR states, and obtain fidelity of 0.9852(6) and 0.9892(9), respectively at a average photon-pair number per pulse $\sim$ 0.01. The Hong-Ou-Mandel measurement with photons from the two independent sources \cite{hong1987measurement} gives a visibility of 0.943(20). We run the experiment and collect 77326 four-photon coincidence detection events over 460810 seconds, i.e., a four-photon coincidence rate of $\sim$0.167 per second. We estimated the three-party correlation for each of Bob's outcome with its related 12 setting combinations and obtain a value of 7.80(6), 7.89(6), 7.78(6), and 7.84(6) for the corresponding Bell states of $\ket{\Phi^{+}}$, $\ket{\Psi^{+}}$, $\ket{\Phi^{-}}$, and $\ket{\Psi^{-}}$, respectively. Summing over all the obtained correlations, we finally get an average result of 7.83(3), which exceeds the real bound of 7.66 over 5.30 standard deviations. The results are described in Fig. \ref{Fig:results}, which provides strong evidence to reject RQM.

We have for the first time demonstrated an experimental refutation of RQM with independent quantum preparations and independent measurement while simultaneously closing locality loopholes under strict locality conditions, which is a significant progress for testing the foundations of quantum mechanics. The current implementation has used two partial BSMs determined by a fast QRNG and random photon clicks to certify the presence of four outcomes. Interestingly, it has been suggested that four outcomes of a full BSM can be reinterpreted as four random setting choices of a QRNG, such that one could use a partial BSM with QRNGs to refute RQM \cite{li2021testing}. Our next step is to close detection loopholes, which appear feasible using high-efficiency photon sources and detectors \cite{rowe2001experimental,larsson2014bell,giustina2013bell}.

\begin{acknowledgements}
$^{\ast}$D.W, Y.-F.J, and X.-M.G contributed equally to this work.

We thank Chang Liu and Quantum Ctek for providing the components used in the QRNGs. This work was supported by the National Natural Science Foundation of China, the Chinese Academy of Sciences, the National Fundamental Research Program, and the Anhui Initiative in Quantum Information Technologies.
\end{acknowledgements}

\bibliographystyle{unsrt}
\bibliography{refs}

\begin{thebibliography}{10}

\bibitem{dirac1981principles}
Paul Adrien~Maurice Dirac.
\newblock {\em The principles of quantum mechanics}.
\newblock Number~27. Oxford university press, 1981.

\bibitem{schrodinger1926undulatory}
Erwin Schr{\"o}dinger.
\newblock An undulatory theory of the mechanics of atoms and molecules.
\newblock {\em Physical review}, 28(6):1049, 1926.

\bibitem{Ismael2021}
J~Ismael.
\newblock in the stanford encyclopedia of philosophy fall 2021 edition (ed.
  zalta, e. n.) (metaphysics research lab, stanford university, 2021).

\bibitem{birkhoff1936logic}
Garrett Birkhoff and John Von~Neumann.
\newblock The logic of quantum mechanics.
\newblock {\em Annals of mathematics}, pages 823--843, 1936.

\bibitem{stueckelberg1959field}
E~CG Stueckelberg.
\newblock Field quantization and time reversal in real hilbert space.
\newblock Technical report, European Organization for Nuclear Research, Geneva,
  1959.

\bibitem{stueckelberg1960quantum}
Ernst~CG Stueckelberg.
\newblock Quantum theory in real hilbert space.
\newblock {\em Helv. Phys. Acta}, 33(727):458, 1960.

\bibitem{guenin1961quantum}
M~Guenin and ECG Stueckelberg.
\newblock Quantum theory in real hilbert space. ii.(addenda and errata).
\newblock {\em Helv. Phys. Acta}, 34:621--628, 1961.

\bibitem{dyson1962threefold}
Freeman~J Dyson.
\newblock The threefold way. algebraic structure of symmetry groups and
  ensembles in quantum mechanics.
\newblock {\em Journal of Mathematical Physics}, 3(6):1199--1215, 1962.

\bibitem{pal2008efficiency}
Karoly~F P{\'a}l and Tam{\'a}s V{\'e}rtesi.
\newblock Efficiency of higher-dimensional hilbert spaces for the violation of
  bell inequalities.
\newblock {\em Physical Review A}, 77(4):042105, 2008.

\bibitem{mckague2009simulating}
Matthew McKague, Michele Mosca, and Nicolas Gisin.
\newblock Simulating quantum systems using real hilbert spaces.
\newblock {\em Physical review letters}, 102(2):020505, 2009.

\bibitem{aleksandrova2013real}
Antoniya Aleksandrova, Victoria Borish, and William~K Wootters.
\newblock Real-vector-space quantum theory with a universal quantum bit.
\newblock {\em Physical Review A}, 87(5):052106, 2013.

\bibitem{bell1964einstein}
John~S Bell.
\newblock On the einstein podolsky rosen paradox.
\newblock {\em Physics Physique Fizika}, 1(3):195, 1964.

\bibitem{renou2021quantum}
Marc-Olivier Renou, David Trillo, Mirjam Weilenmann, Thinh~P Le, Armin
  Tavakoli, Nicolas Gisin, Antonio Ac{\'\i}n, and Miguel Navascu{\'e}s.
\newblock Quantum theory based on real numbers can be experimentally falsified.
\newblock {\em Nature}, 600(7890):625--629, 2021.

\bibitem{roos1970independence}
Hansj{\"o}rg Roos.
\newblock Independence of local algebras in quantum field theory.
\newblock {\em Communications in Mathematical Physics}, 16(3):238--246, 1970.

\bibitem{werner1987local}
Reinhard Werner.
\newblock Local preparability of states and the split property in quantum field
  theory.
\newblock {\em Letters in Mathematical Physics}, 13(4):325--329, 1987.

\bibitem{weihs1998violation}
Gregor Weihs, Thomas Jennewein, Christoph Simon, Harald Weinfurter, and Anton
  Zeilinger.
\newblock Violation of bell's inequality under strict einstein locality
  conditions.
\newblock {\em Physical Review Letters}, 81(23):5039, 1998.

\bibitem{scheidl2010violation}
Thomas Scheidl, Rupert Ursin, Johannes Kofler, Sven Ramelow, Xiao-Song Ma,
  Thomas Herbst, Lothar Ratschbacher, Alessandro Fedrizzi, Nathan~K Langford,
  Thomas Jennewein, et~al.
\newblock Violation of local realism with freedom of choice.
\newblock {\em Proceedings of the National Academy of Sciences},
  107(46):19708--19713, 2010.

\bibitem{rowe2001experimental}
Mary~A Rowe, David Kielpinski, Volker Meyer, Charles~A Sackett, Wayne~M Itano,
  Christopher Monroe, and David~J Wineland.
\newblock Experimental violation of a bell's inequality with efficient
  detection.
\newblock {\em Nature}, 409(6822):791--794, 2001.

\bibitem{larsson2014bell}
Jan-{\AA}ke Larsson, Marissa Giustina, Johannes Kofler, Bernhard Wittmann,
  Rupert Ursin, and Sven Ramelow.
\newblock Bell-inequality violation with entangled photons, free of the
  coincidence-time loophole.
\newblock {\em Physical Review A}, 90(3):032107, 2014.

\bibitem{giustina2013bell}
Marissa Giustina, Alexandra Mech, Sven Ramelow, Bernhard Wittmann, Johannes
  Kofler, J{\"o}rn Beyer, Adriana Lita, Brice Calkins, Thomas Gerrits, Sae~Woo
  Nam, et~al.
\newblock Bell violation using entangled photons without the fair-sampling
  assumption.
\newblock {\em Nature}, 497(7448):227--230, 2013.

\bibitem{giustina2015significant}
Marissa Giustina, Marijn~AM Versteegh, S{\"o}ren Wengerowsky, Johannes
  Handsteiner, Armin Hochrainer, Kevin Phelan, Fabian Steinlechner, Johannes
  Kofler, Jan-{\AA}ke Larsson, Carlos Abell{\'a}n, et~al.
\newblock Significant-loophole-free test of bell's theorem with entangled
  photons.
\newblock {\em Physical review letters}, 115(25):250401, 2015.

\bibitem{hensen2015loophole}
Bas Hensen, Hannes Bernien, Ana{\"\i}s~E Dr{\'e}au, Andreas Reiserer, Norbert
  Kalb, Machiel~S Blok, Just Ruitenberg, Raymond~FL Vermeulen, Raymond~N
  Schouten, Carlos Abell{\'a}n, et~al.
\newblock Loophole-free bell inequality violation using electron spins
  separated by 1.3 kilometres.
\newblock {\em Nature}, 526(7575):682--686, 2015.

\bibitem{shalm2015strong}
Lynden~K Shalm, Evan Meyer-Scott, Bradley~G Christensen, Peter Bierhorst,
  Michael~A Wayne, Martin~J Stevens, Thomas Gerrits, Scott Glancy, Deny~R
  Hamel, Michael~S Allman, et~al.
\newblock Strong loophole-free test of local realism.
\newblock {\em Physical review letters}, 115(25):250402, 2015.

\bibitem{rosenfeld2017event}
Wenjamin Rosenfeld, Daniel Burchardt, Robert Garthoff, Kai Redeker, Norbert
  Ortegel, Markus Rau, and Harald Weinfurter.
\newblock Event-ready bell test using entangled atoms simultaneously closing
  detection and locality loopholes.
\newblock {\em Physical review letters}, 119(1):010402, 2017.

\bibitem{li2018test}
Ming-Han Li, Cheng Wu, Yanbao Zhang, Wen-Zhao Liu, Bing Bai, Yang Liu, Weijun
  Zhang, Qi~Zhao, Hao Li, Zhen Wang, et~al.
\newblock Test of local realism into the past without detection and locality
  loopholes.
\newblock {\em Physical review letters}, 121(8):080404, 2018.

\bibitem{chen2021ruling}
Ming-Cheng Chen, Can Wang, Feng-Ming Liu, Jian-Wen Wang, Chong Ying, Zhong-Xia
  Shang, Yulin Wu, M~Gong, H~Deng, F-T Liang, et~al.
\newblock Ruling out real-valued standard formalism of quantum theory.
\newblock {\em Physical Review Letters}, 128(4):040403, 2022.

\bibitem{barends2013coherent}
Rami Barends, Julian Kelly, Anthony Megrant, Daniel Sank, Evan Jeffrey,
  Yu~Chen, Yi~Yin, Ben Chiaro, Josh Mutus, Charles Neill, et~al.
\newblock Coherent josephson qubit suitable for scalable quantum integrated
  circuits.
\newblock {\em Physical review letters}, 111(8):080502, 2013.

\bibitem{li2021testing}
Zheng-Da Li, Ya-Li Mao, Mirjam Weilenmann, Armin Tavakoli, Hu~Chen, Lixin Feng,
  Sheng-Jun Yang, Marc-Olivier Renou, David Trillo, Thinh~P Le, et~al.
\newblock Testing real quantum theory in an optical quantum network.
\newblock {\em Physical Review Letters}, 128(4):040402, 2022.

\bibitem{qi2010high}
Bing Qi, Yue-Meng Chi, Hoi-Kwong Lo, and Li~Qian.
\newblock High-speed quantum random number generation by measuring phase noise
  of a single-mode laser.
\newblock {\em Optics letters}, 35(3):312--314, 2010.

\bibitem{nie2015generation}
You-Qi Nie, Leilei Huang, Yang Liu, Frank Payne, Jun Zhang, and Jian-Wei Pan.
\newblock The generation of 68 gbps quantum random number by measuring laser
  phase fluctuations.
\newblock {\em Review of Scientific Instruments}, 86(6):063105, 2015.

\bibitem{einstein1935can}
Albert Einstein, Boris Podolsky, and Nathan Rosen.
\newblock Can quantum-mechanical description of physical reality be considered
  complete?
\newblock {\em Physical review}, 47(10):777, 1935.

\bibitem{sun2019experimental}
Qi-Chao Sun, Yang-Fan Jiang, Bing Bai, Weijun Zhang, Hao Li, Xiao Jiang, Jun
  Zhang, Lixing You, Xianfeng Chen, Zhen Wang, et~al.
\newblock Experimental demonstration of non-bilocality with truly independent
  sources and strict locality constraints.
\newblock {\em Nature Photonics}, 13(10):687--691, 2019.

\bibitem{weinfurter1994experimental}
Harald Weinfurter.
\newblock Experimental bell-state analysis.
\newblock {\em EPL (Europhysics Letters)}, 25(8):559, 1994.

\bibitem{calsamiglia2001maximum}
John Calsamiglia and Norbert L{\"u}tkenhaus.
\newblock Maximum efficiency of a linear-optical bell-state analyzer.
\newblock {\em Applied Physics B}, 72(1):67--71, 2001.

\bibitem{brunner2014bell}
Nicolas Brunner, Daniel Cavalcanti, Stefano Pironio, Valerio Scarani, and
  Stephanie Wehner.
\newblock Bell nonlocality.
\newblock {\em Reviews of Modern Physics}, 86(2):419, 2014.

\bibitem{hong1987measurement}
Chong-Ki Hong, Zhe-Yu Ou, and Leonard Mandel.
\newblock Measurement of subpicosecond time intervals between two photons by
  interference.
\newblock {\em Physical review letters}, 59(18):2044, 1987.

\bibitem{rukhin2010statistical}
A~Rukhin, J~Soto, J~Nechvatal, M~Smid, E~Barker, S~Leigh, M~Levenson, M~Vangel,
  D~Banks, A~Heckert, et~al.
\newblock A statistical test suite for random and pseudorandom number
  generators for cryptographic applications-special publication 800-22 rev1a.
\newblock {\em Tech. Rep. April}, 2010.

\end{thebibliography}
 \newpage
 \onecolumngrid
\begin{center}
 \Large Supplementary Information for Experimental refutation of real quantum mechanics under strict locality conditions\\
\end{center}

 \normalsize
 \subsection{1. Polarization entangled photon-pair sources and Hong-Ou-Mandel interference}
\begin{figure}[!h]
 	\centering
 	\includegraphics[width=0.85\textwidth]{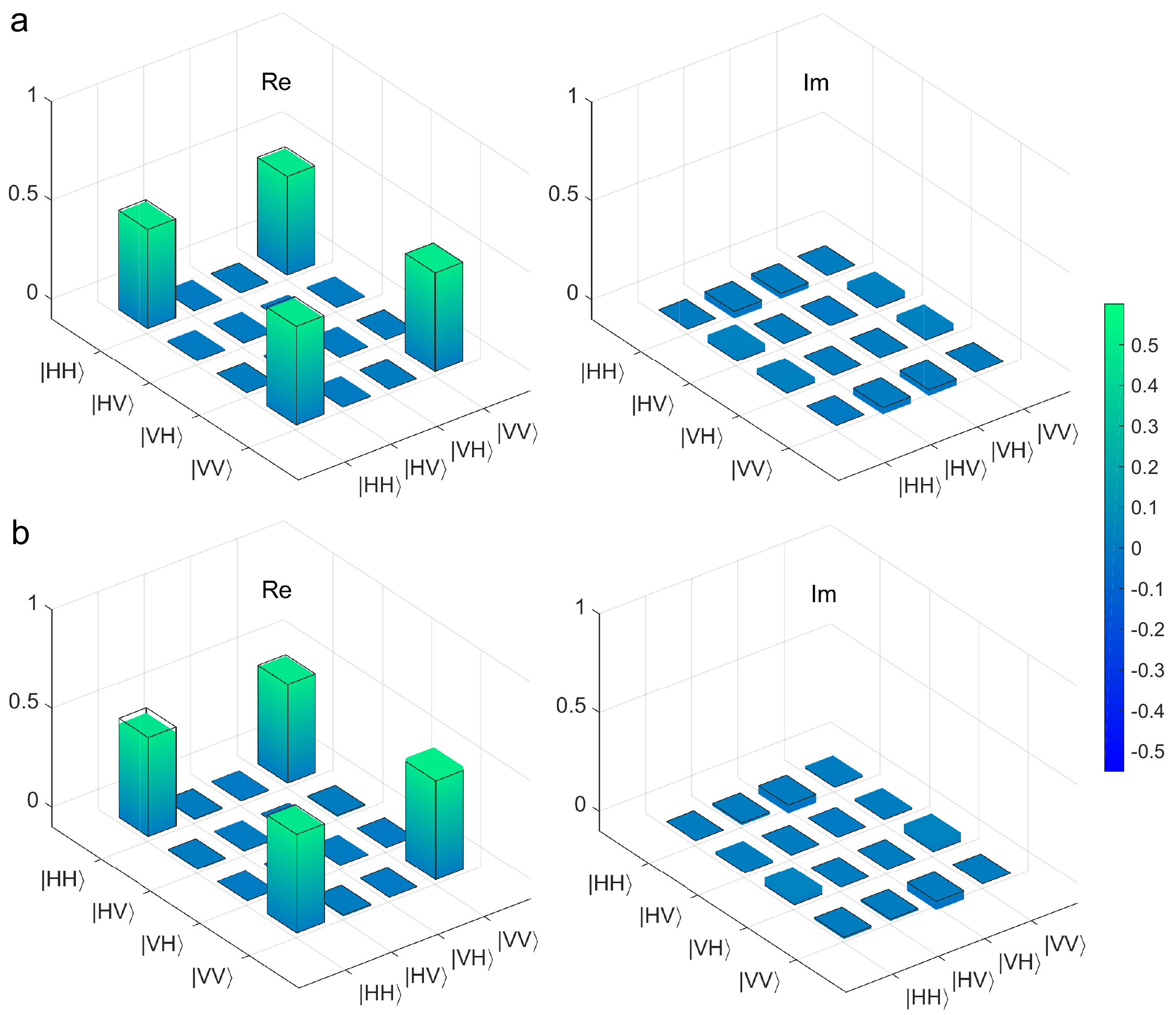}
 	\caption{The result of quantum state tomography measurement. \textbf{a}, real part and imaginary part of the reconstructed density matrix for source $\text{S}_{1}$. Black frames indicate the density matrix of ideal EPR state $|\Phi \rangle$. \textbf{b}, real part and imaginary part of the reconstructed density matrix for source $\text{S}_{2}$.} \label{Fig:Tomoresult}
 \end{figure}

\begin{figure}[!t]
 	\centering
	\includegraphics[width=0.58\textwidth]{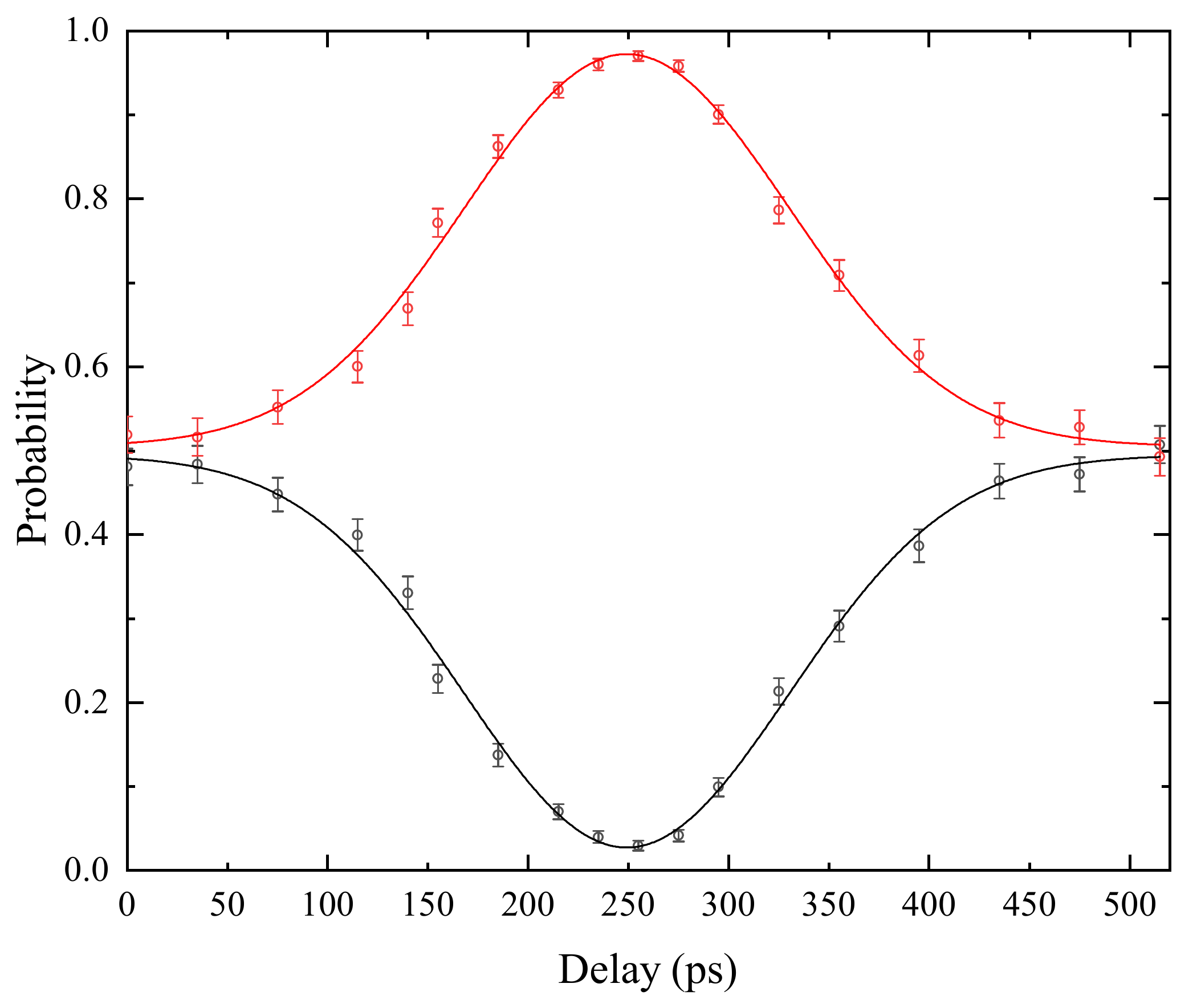}
	\caption{Experiment result of two-photon Hong-Ou-Mandel interference. Each data point is accumulated for more than 600s. The fitted visibility is $0.943 \pm 0.020$}\label{Fig:HOMresult}
 \end{figure}

 In the main text, we generate polarization entangled photon-pairs in the Bell state $\ket{\Phi^{+}}=(\ket{HH}+\ket{VV})/\sqrt{2}$ via the Type-0 spontaneous parametric down conversion (SPDC) process in the PPMgLN crystal, where $\ket{H}$ and $\ket{V}$ are the horizontal and vertical polarization, respectively. To suppress the distinguishability between photons from the two separate sources in our experiment, we pass photons through inline 3.3 GHz fibre Bragg gratings (FBG) to suppress the spectral distinguishability. The 133 ps coherence time of single photons is much longer than the pump pulse duration, which, together with the high bandwidth synchronization (with an uncertainty of 4 ps), suppress the temporal distinguishability; the good fibre optical mode eliminates the spatial distinguishability \cite{sun2019experimental}. We then reconstruct the density matrices of the two sources $\text{S}_{1}$ and $\text{S}_{2}$ based on quantum state tomography measurement respectively with fidelities $0.9852\pm0.0006 $ and $0.9892\pm 0.0009 $ with respect to the ideal state $|\Phi^+\rangle$, as depicted in Fig. \ref{Fig:Tomoresult}. The uncertainties are obtained using a Monte Carlo routine assuming Poissonian statistics. Shown in Fig. \ref{Fig:HOMresult} is the Hong-Ou-Mandel measure by Bob with photons from the two independent sources, and we obtain a fitted visibility of $0.943\pm0.020$.
\subsection{2. Quantum random number generators, polarization measurement setting, and experimental results}
To disproof the real number quantum mechanics, Alice randomly selects one of the six measurement settings $A_{x}\in\{Z\pm X, Z\pm Y, X\pm Y\}/\sqrt{2}$, and Claire randomly chose one of three measurement settings $C_{z}\in\{Z,X,Y\}$. Bob performs a full Bell-state measurement deiced by real-time quantum random number generator (QRNG) and random photon clicks. All single photon polarization state modulations are controlled by random inputs of the real-time QRNGs at a rate of 250 MHz.

The QRNGs used in our experiment are based on the quantum phase noise of a single-mode laser. A 1550-nm laser diode is driven by a constant current which is near its threshold, and a thermo electric cooler is used to stabilize its temperature. The emitted photons enter an unbalanced interferometer to convert the phase information into the intensity. The output of the interferometer is detected by a 10-GHz InGaAs photon detector (PD). The PD signals is converted to 8 bits per sample by an analog-to-digital converter (ADC) at a repetition rate of 1 GHz. We use 1- or 2-bits random numbers per duty to satisfy all random measurement basis choice. We choose one or two bits per sample and feed them into a field-programmable gate array (FPGA) respectively. In order to obtain better randomness distribution and reduce data processing delay, we use the XOR algorithm which consumes every 4 adjacent bits to generate 1 random bit for the measurement setting choice at a repetition rate of 250 MHz. The post-processing is performed by the FPGA in real time. The time elapse that QRNGs take to generate a bit used in our experiment is defined as the time interval between the start of the experiment trail and the FPGA output an electrical signal. When QRNG outputs 1 bit per duty, the calibrated time interval is less than 35.5 ns. Meanwhile with 2 bits output per duty, we set longer interval time to reduce the data autocorrelation brought by the same random source. We finally added an extra time with 36 ns between two generated bits. All the generated random bits pass the NIST statistical test suite \cite{rukhin2010statistical}, as shown in Table \ref{Nist tests}.

\begin{figure*}[!h]
 	\centering
 	\includegraphics[width=1\textwidth]{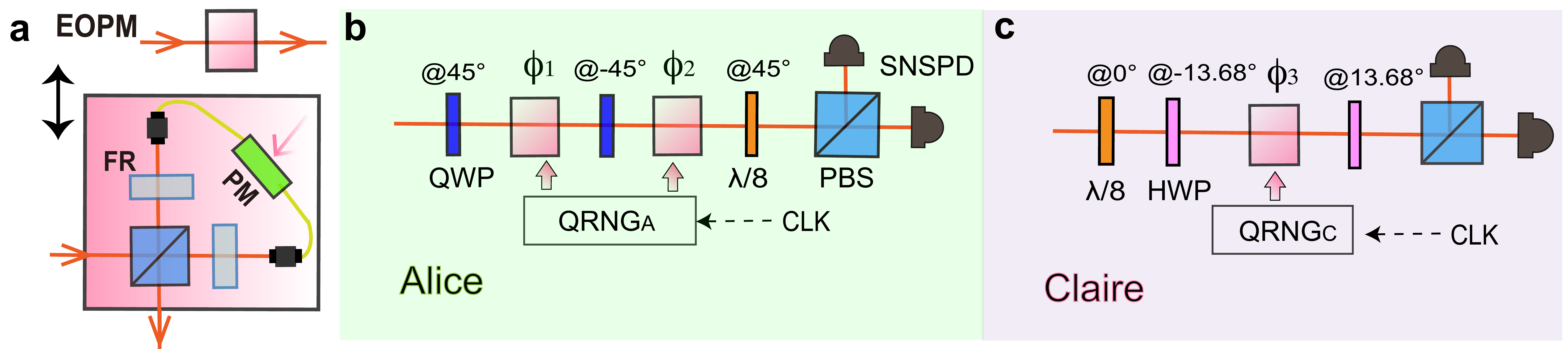}
 	\caption{The setup of random measurement basis choice. \textbf{a}, high-speed electro-optical phase modulation loop; \textbf{b}, Alice performs one of six polarization measurement bases ($A_{x}\in\{Z\pm X, Z\pm Y, X\pm Y\}/\sqrt{2}$) and the setting choice is determined by the co-located ($\text{QRNG}_{A}$). \textbf{c}, Similar to \textbf{b},  Claire performs her measurement setting in one of three bases ($C_{z}\in\{Z,X,Y\}$).}
 	\label{Fig:Settings}
 \end{figure*}

In this experiment, we use the high-speed electro-optical phase modulation loops (EOPMs, shown in Fig. \ref{Fig:Settings}a) and specific waveplates to realize the single photon polarization state modulation at a rate of 250 MHz (Fig. \ref{Fig:Settings}). The EOPM consists of a polarization beam splitter (PBS), two Faraday rotators (FR) and a fiber optic phase modulator (PM): when a single photon enters the EOPM, its two orthogonal polarization components exit at different ports of the PBS, both of the two components are aligned to slow axis of the FPM after rotating 45 degrees via FR. The fiber-based PM is displaced 20 cm from the middle position such that the two polarization components arrive at the PM with a time difference of 2 ns. Hence, we can attach a phase to one of the polarizations alone. The two polarization components merge at the PBS and exit in a modulated polarization state. The matrix of an EOPM can be written as $P\left(\phi\right)=\left(\begin{matrix}1&0\\0&e^{i\phi}\\\end{matrix}\right)$, where $\phi$ is controlled by QRNGs. The detailed configuration of the measurement settings at Alice's and Claire's station are shown in Table \ref{tab:alicesetting} (a) and \ref{tab:clairesetting} (b). The phase $\phi_1$ is controlled by 1-bit QRNG while $\phi_2$ and $\phi_3$ are triggered by 2-bit QRNG outputs. Note that, we only need 3 different random outputs, thus the redundant outputs are discarded. We defined $F_m=\frac{C_r}{C_r+C_w}$ as the fidelity of random basis choice, where $C_r$ represents the photons recorded by the correct superconducting nanowire single photon detectors (SNSPDs) and $C_w$ the photons recorded by wrong SNSPDs, respectively.
 \begin{table*}[!h]
 	\caption{Results of the NIST test suite after dividing 1Gbit QRNG outputs into 1 Mbit sections}
 	\begin{tabular}{cccccccccc}
 		\hline
 		\multicolumn{1}{c|}{Location}     & \multicolumn{3}{c|}{Alice}                                                      & \multicolumn{3}{c|}{Bob}                                                        & \multicolumn{3}{c}{Claire}                                \\ \hline
 		\multicolumn{1}{c|}{Statistical tests}       & P value            & Proportion           & \multicolumn{1}{c|}{Result}       & P value           & Proportion           & \multicolumn{1}{c|}{Result}       & P value             & Proportion           & Result       \\ \hline
 		\multicolumn{1}{c|}{Frequency}               & 0.197981             & 0.989                & \multicolumn{1}{c|}{$\checkmark$} & 0.03732              & 0.990                & \multicolumn{1}{c|}{$\checkmark$} & 0.542228             & 0.985                & $\checkmark$ \\
 		\multicolumn{1}{c|}{BlockFrequency}          & 0.127393             & 0.993                & \multicolumn{1}{c|}{$\checkmark$} & 0.872425             & 0.990                 & \multicolumn{1}{c|}{$\checkmark$} & 0.903338             & 0.986                & $\checkmark$ \\
 		\multicolumn{1}{c|}{CumulativeSums}          & 0.217714             & 0.987                & \multicolumn{1}{c|}{$\checkmark$} & 0.462924             & 0.990               & \multicolumn{1}{c|}{$\checkmark$} & 0.284927             & 0.985               & $\checkmark$ \\
 		\multicolumn{1}{c|}{Runs}                    & 0.603841             & 0.991                & \multicolumn{1}{c|}{$\checkmark$} & 0.15991              & 0.991                & \multicolumn{1}{c|}{$\checkmark$} & 0.97748              & 0.988                & $\checkmark$ \\
 		\multicolumn{1}{c|}{LongestRun}              & 0.872425             & 0.994                & \multicolumn{1}{c|}{$\checkmark$} & 0.075719             & 0.989                & \multicolumn{1}{c|}{$\checkmark$} & 0.78683              & 0.986                & $\checkmark$ \\
 		\multicolumn{1}{c|}{Rank}                    & 0.765632             & 0.990                 & \multicolumn{1}{c|}{$\checkmark$} & 0.800005             & 0.991                & \multicolumn{1}{c|}{$\checkmark$} & 0.666245             & 0.992                & $\checkmark$ \\
 		\multicolumn{1}{c|}{FFT}                     & 0.006519             & 0.986                & \multicolumn{1}{c|}{$\checkmark$} & 0.737915             & 0.990                & \multicolumn{1}{c|}{$\checkmark$} & 0.420827             & 0.990                & $\checkmark$ \\
 		\multicolumn{1}{c|}{NonOverlappingTemplate}  & 0.080142             & 0.990             & \multicolumn{1}{c|}{$\checkmark$} & 0.072858             & 0.990             & \multicolumn{1}{c|}{$\checkmark$} & 0.057098405          & 0.990          & $\checkmark$ \\
 		\multicolumn{1}{c|}{OverlappingTemplate}     & 0.080519             & 0.987                & \multicolumn{1}{c|}{$\checkmark$} & 0.049346             & 0.985                & \multicolumn{1}{c|}{$\checkmark$} & 0.618385             & 0.989                & $\checkmark$ \\
 		\multicolumn{1}{c|}{Universal}               & 0.446556             & 0.985                & \multicolumn{1}{c|}{$\checkmark$} & 0.087162             & 0.989                & \multicolumn{1}{c|}{$\checkmark$} & 0.030197             & 0.988                & $\checkmark$ \\
 		\multicolumn{1}{c|}{ApproximateEntropy}      & 0.747898             & 0.990                & \multicolumn{1}{c|}{$\checkmark$} & 0.798139             & 0.993                & \multicolumn{1}{c|}{$\checkmark$} & 0.925287             & 0.990                & $\checkmark$ \\
 		\multicolumn{1}{c|}{RandomExcursions}        & 0.241294             & 0.989             & \multicolumn{1}{c|}{$\checkmark$} & 0.197918             & 0.988             & \multicolumn{1}{c|}{$\checkmark$} & 0.419019             & 0.989          & $\checkmark$ \\
 		\multicolumn{1}{c|}{RandomExcursionsVariant} & 0.19579              & 0.990             & \multicolumn{1}{c|}{$\checkmark$} & 0.155235             & 0.988             & \multicolumn{1}{c|}{$\checkmark$} & 0.188237444          &0.991          & $\checkmark$ \\
 		\multicolumn{1}{c|}{Serial}                  & 0.211272             & 0.989               & \multicolumn{1}{c|}{$\checkmark$} & 0.15119              & 0.990               & \multicolumn{1}{c|}{$\checkmark$} & 0.383695             &0.994               & $\checkmark$ \\
 		\multicolumn{1}{c|}{LinearComplexity}        & 0.769527             & 0.995                & \multicolumn{1}{c|}{$\checkmark$} & 0.478839             & 0.991                & \multicolumn{1}{c|}{$\checkmark$} & 0.293952             & 0.991                & $\checkmark$ \\          
 	\end{tabular}
 	\label{Nist tests}
 \end{table*}

 \begin{table}[!h]
 	\centering
 	\caption{Alice's and Claire's measurements setting and fidelity}
 	\renewcommand{\arraystretch}{1.5}
 	\begin{minipage}{.5\linewidth}
 		\begin{tabular}{|>{\centering}m{1.5em}|>{\centering}m{6em}|>{\centering}m{3.em}|>{\centering}m{2em}|>{\centering}m{3.em}|>{\centering}m{2.em}|>{\centering}m{3em}|c|>{\centering}m{2em}|}
 			\hline
 			$x$  &  $A_x$& QWP& $\phi_1$ & QWP& $\phi_2$ &$\lambda$/8   &$F_{Ax}$ \\ \hline
 			1  & $(Z+X)/\sqrt{2}$ & @45$^{\circ}$& 0     &@-45$^{\circ}$& $\frac{\pi}{2}$ &@45$^{\circ}$& 0.996    \\ \hline
 			2  & $(Z-X)/\sqrt{2}$  &@45$^{\circ}$ &$\frac{\pi}{2}$& @-45$^{\circ}$& $\frac{\pi}{2}$ &@45$^{\circ}$ & 0.993   \\ \hline
 			3  & $(Z+Y)/\sqrt{2}$ & @45$^{\circ}$&0              & @-45$^{\circ}$& 0               &@45$^{\circ}$& 0.994    \\ \hline
 			4  & $(Z-Y)/\sqrt{2}$  & @45$^{\circ}$&0              & @-45$^{\circ}$& $\pi$          & @45$^{\circ}$& 0.995    \\ \hline
 			5  & $(X+X)/\sqrt{2}$ & @45$^{\circ}$&$\frac{\pi}{2}$&@-45$^{\circ}$ & $\pi$         & @45$^{\circ}$& 0.992    \\ \hline
 			6  & $(X-Y)/\sqrt{2}$ &@45$^{\circ}$ &$\frac{\pi}{2}$ &@-45$^{\circ}$& 0              & @45$^{\circ}$& 0.993    \\ \hline
 		\end{tabular}
 			\vspace{0.1cm} 
 		\\(a) Alice's six measurement settings $A_{x}$ decided by three inputs $x$ of $\text{QRNG}_{A}$
 		\label{tab:alicesetting}
 	\end{minipage}%
 	\hfill
 	\begin{minipage}{.5\linewidth}
 		\begin{tabular}{|>{\centering}m{1.5em}|>{\centering}m{2.em}|>{\centering}m{2em}|>{\centering}m{4em}|>{\centering}m{2em}|>{\centering}m{4.em}|c|>{\centering}m{3.em}|}
 			\hline
 			$z$ & $C_z$ & $\lambda$/8& HWP &  $\phi_3$&  HWP& $F_{Cz}$ \\ \hline
 			1 & $Z$  &@0$^{\circ}$& @-13.68$^{\circ}$& 0           & @13.68$^{\circ}$& 0.994    \\ \hline
 			2 & $X$   &@0$^{\circ}$&@-13.68$^{\circ}$& $\frac{2\pi}{3}$ &@13.68$^{\circ}$& 0.995    \\ \hline
 			3 & $Y$  &@0$^{\circ}$&@-13.68$^{\circ}$ & $\frac{4\pi}{3}$ &@13.68$^{\circ}$& 0.991    \\ \hline
 		\end{tabular}
 		\vspace{0.1cm} 
 		\\(b) Claire's three measurement settings $C_{z}$ decided by three inputs $z$ of $\text{QRNG}_{c}$
 		\label{tab:clairesetting}
 	\end{minipage}%
 \end{table}

\begin{figure*}[!h]
 	\centering
 	\includegraphics[width=1\textwidth]{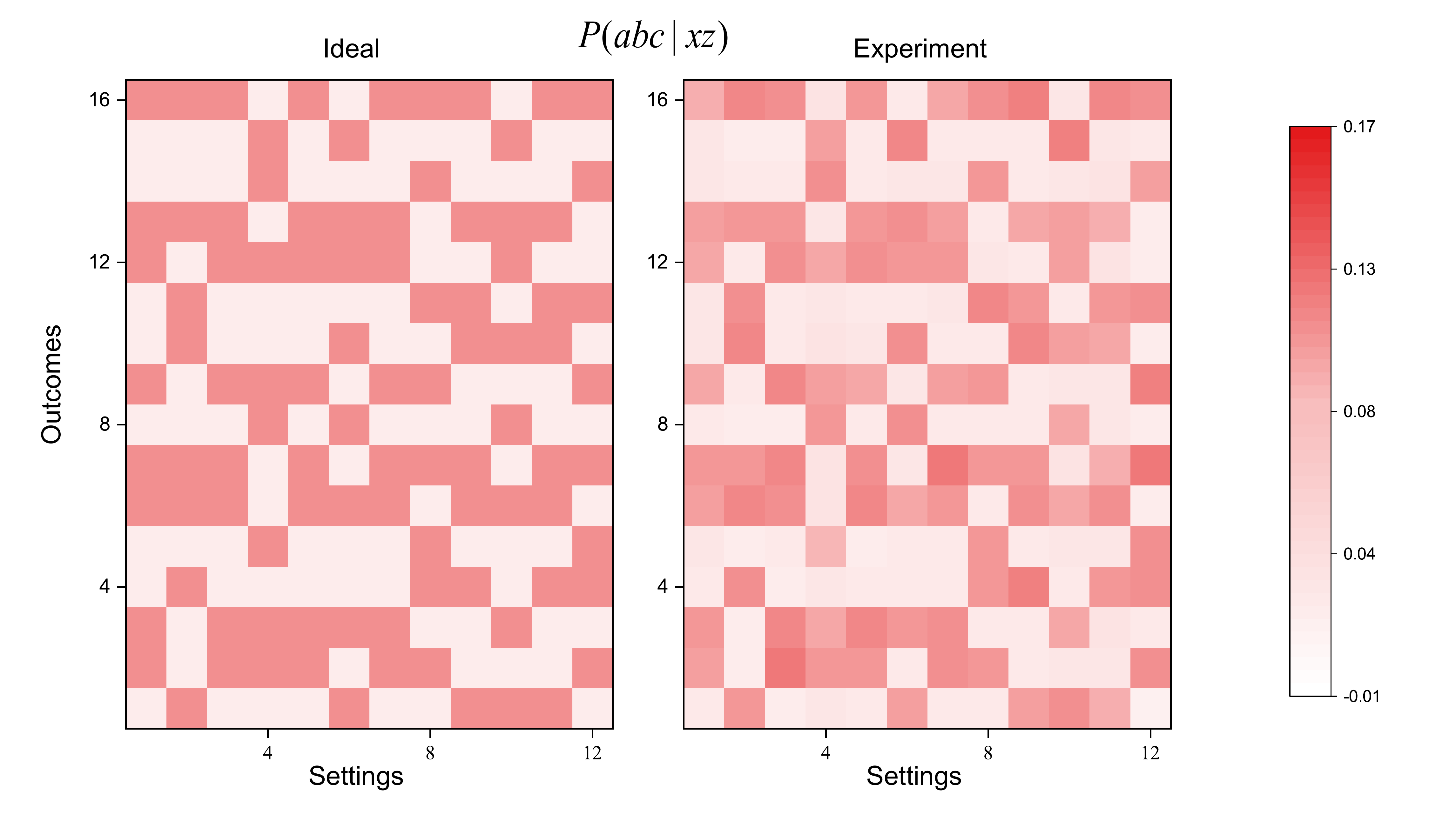}
 	\caption{Experiment result. The conditional joint probability distribution matrices $\text{P}(abc|xz)$ from the non-local game (left) and the theoretical prediction from an ideal experiment (right). The row index represents the measurement setting ${xz}$ and the column index is the measured strings ${abc}$.}
 	\label{Fig:Presults}
 \end{figure*}

As shown in the main text, Alice and Claire independently perform their measurement $A_{x}$ and $C_{z}$ with random inputs $x\in\{1,2,3,4,5,6\}$ and $z\in\{1,2,3\}$, and produce outcomes as $a,c=\pm1$. Bob performs a single measurement labeled as input $y$ that returns four outcomes $b=b_{1}b_{2}\in\{0,1\}^2$. The three-party correlation is defined as the weighted sum of a input-output probability distribution $p(abc|xz)$ \cite{renou2021quantum}, given by $\mathcal{F}=\sum_{abc,xz}^{}\omega_{abc,xz} p(abc|xz)$, where $w_{abc,xz}=\pm1$ are the weights, $abc\in\{0,1\}^{\otimes^{4}}$ are the bit strings of the measurement results. There are 12 different combinations of measurement settings $xz\in\{11,12,21,22,13,14,33,34,52,53,62,63\}$ and 16 possible measured bit strings. The experimental results are plotted as a 12$\times$16 probability distribution matrix in Fig. \ref{Fig:Presults} together with the ideal values.

 \subsection{3. Synchronization of the setup} 
\begin{figure*}[!h]
 	\centering
 	\includegraphics[width=1\textwidth]{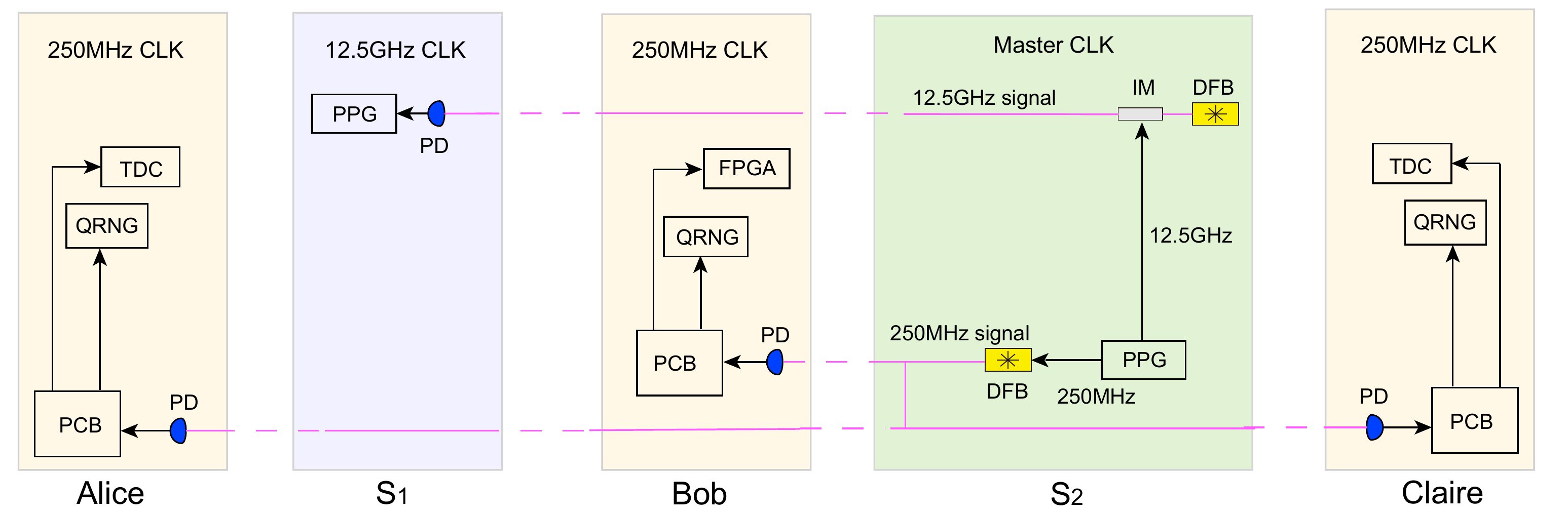}
 	\caption{Schematics for the experimental synchronization. The red dashed lines and black lines represent the fiber channels and coaxial cables, respectively.} \label{Fig:Time sync}
 \end{figure*}

The pulse pattern generator (PPG) located at Source $\text{S}_{2}$ acts as a master clock, as shown in Fig. \ref{Fig:Time sync}. It generates a 12.5 GHz sinusoidal signal and a 250 MHz square wave signal to prepare the optical signal for synchronization, respectively. After electro-optical conversion, the 12.5 GHz optical signal is used to synchronize the PPG located at $\text{S}_{1}$ , and the 250 MHz optical signals are used to synchronize all other devices.
 \subsection{4. Space-time analysis of the experiment}
 Strict locality constraints should be met in our experiment in order to close the locality loophole and ensure independent state preparations and measurements. The relative positions of the two sources and three measurement stations in our experiment are shown in Table. \ref{tab:spacedistance} (a) and the length of fiber links between them are described in Table. \ref{tab:fiberdistance} (b). The locality constraints set seven space-like separation conditions in each experimental trial as follows:
 
 (1). Space-like separation between the two state emission events in sources $\text{S}_{1}$ and $\text{S}_{2}$ requires $\text{L}_{12}/c>\text{max}(t_{12}+\tau_{1}, t_{12}+\tau_2)$, where $\text{L}_{12}$ is the beeline distance between $\text{S}_{1}$ and $\text{S}_{2}$, $t_{12}$ is the relative delay between the earliest time of the state emission (the creation of pump laser pulses in the two sources), which are also taken as the earliest time to create the local hidden variables, $\tau_1 (\tau_2)$ is the time elapse starting from the earliest time of generating a pump laser pulse to the latest time of loading the photon to the optical fiber for entanglement distribution.

 (2). Space-like separation between the event of generating quantum random numbers for Alice ($\text{QRNG}_{A}$) and the event of generating quantum random numbers for Claire ($\text{QRNG}_{C}$) requires $\text{L}_{AC}/c> \text{max}(t_{AqCq}+\tau_{Aq}, t_{AqCq}+\tau_{Cq})$, where $\text{L}_{AC}$ is the distance between node Alice and node Claire, and $t_{AqCq}$ is the relative delay between the earliest time of $\text{QRNG}_{A}$ and that of $\text{QRNG}_{C}$, $\tau_{Aq}$ is the time elapse for $\text{QRNG}_{A}$ to output three bits randomly, and $\tau_{Cq}$ is the time elapse for $\text{QRNG}_{C}$ to output two bits randomly.
 
 (3). Space-like separation between $\text{QRNG}_{A}$ and the state emission events in sources $\text{S}_{1}$ and $\text{S}_{2}$, respectively requires $\text{L}_{A1}/c>(t_{Aq1}+\tau_{Aq})$ and $\text{L}_{A2}/c>(t_{Aq2}+\tau_{Aq})$, where $\text{L}_{A1}$ ($\text{L}_{A2}$) is the beeline distance between Alice and source $\text{S}_{1}$ ($\text{S}_{2}$), $t_{Aq1}$ ($t_{Aq2}$) denotes the relative delay between the earliest time of quantum random number generation and the earliest time of state emission in sources. Similarly, space-like separation between $\text{QRNG}_{c}$ and the state emission events in sources $\text{S}_{1}$ and $\text{S}_{2}$, respectively requires $\text{L}_{C1}/c>(t_{Cq1}+\tau_{Cq})$ and $\text{L}_{C2}/c>(t_{Cq2}+\tau_{Cq})$, where $\text{L}_{C1}$ ($\text{L}_{C2}$) is the spatial distance between Alice and source $\text{S}_{1}$ ($\text{S}_{2}$), $t_{Cq1}$ ($t_{Cq2}$) is the relative delay between the earliest time of quantum random number generation and the earliest time of state emission in sources.

 (4). Space-like separation between $\text{QRNG}_{A}$ and the measurement events by Bob ($\text{M}_{B}$) and Claire ($\text{M}_{C}$) requires $\text{L}_{AB}/c> t_{AqB}+\tau_B$ and $\text{L}_{AC}/c> t_{AqC}+\tau_C$, where $L_{AB}$ and $L_{AC}$ are the beeline distance between the nodes, $t_{AqB}$ ($t_{AqC}$) stands for the relative delay between the earliest time of $\text{QRNG}_{A}$ and the measurement events $\text{M}_{B}$ ($\text{M}_{C}$), and $\tau_{B}$ ($\tau_{C}$) is the time elapse for the measurement event $\text{M}_{B}$ ($\text{M}_{C}$), which is the interval between the time when a photon enters the loop interferometer for polarization measurement and the time when the SNSPD outputs a signal. Similarly, space-like separation between $\text{QRNG}_{C}$ and the measurement events by Alice $\text{M}_{A}$ and Bob requires $\text{L}_{CA}/c> t_{CqA}+\tau_A$ and $\text{L}_{CB}/c> t_{CqB}+\tau_B$, where $\text{L}_{CA}$ and $\text{L}_{CB}$ denote the beeline distance between the nodes, $t_{CqA}$ ($t_{CqB}$) denotes the relative delay between the earliest time of $\text{QRNG}_{C}$ and $\text{M}_{A}$ ($\text{M}_{B}$), and  $\tau_{A}$ is the time elapse for the measurement events, which is also the interval between the time when a photon enters the loop interferometer for polarization measurement and the time when the SNSPD outputs a signal.
 
 (5). Space-like separation between the event of generating quantum random numbers for Bob $\text{QRNG}_{B}$ and the state emission events in sources $\text{S}_{1}$ and $\text{S}_{2}$, respectively requires $\text{L}_{B1}/c>(t_{Bq1}+\tau_{Bq})$ and $\text{L}_{B2}/c>(t_{Bq2}+\tau_{Bq})$, where $\text{L}_{B1}$ ($\text{L}_{B2}$) is the beeline distance between Bob and source $\text{S}_{1}$ ($\text{S}_{2}$), $t_{Bq1}$ ($t_{Bq2}$) denotes the relative delay between the earliest time of quantum random number generation and the earliest time of state emission in source $\text{S}_{1}$ ($\text{S}_{2}$).

 (6). Space-like separation between the event of generating quantum random numbers for Bob $\text{QRNG}_{B}$ and that of $\text{QRNG}_{A}$ requires $\text{L}_{BA}/c> \text{max}(t_{BqAq}+\tau_{Bq}, t_{BqAq}+\tau_{Aq})$, where $\text{L}_{BA}$ is the distance between node Bob and node Alice, and $t_{BqAq}$ is the relative delay between the earliest time of $\text{QRNG}_{B}$ and that of $\text{QRNG}_{A}$, $\tau_{Bq}$ is the time elapse for $\text{QRNG}_{B}$ to output a random bit. Similarly, space-like separation between $\text{QRNG}_{B}$ and event $\text{QRNG}_{C}$ requires $\text{L}_{BC}/c> \text{max}(t_{BqCq}+\tau_{Bq}, t_{BqCq}+\tau_{Cq})$, where $\text{L}_{BC}$ is the distance between node Bob and node Claire, and $t_{BqCq}$ is the relative delay between the earliest time of $\text{QRNG}_{B}$ and that of $\text{QRNG}_{C}$.

 (7). Space-like separation between $\text{QRNG}_{B}$ and the measurement events by Alice ($\text{M}_{A}$) and Claire ($\text{M}_{C}$) requires $\text{L}_{BA}/c> t_{BqA}+\tau_A$ and $\text{L}_{BC}/c> t_{BqC}+\tau_C$, where $L_{BA}$ and $L_{BC}$ are the beeline distance between the nodes, $t_{BqA}$ ($t_{BqC}$) stands for the relative delay between the earliest time of $\text{QRNG}_{B}$ and the measurement events $\text{M}_{A}$ ($\text{M}_{C}$), and $\tau_{A}$ ($\tau_{C}$) is the time elapse for the measurement event $\text{M}_{A}$ ($\text{M}_{C}$), which is the interval between the time when a photon enters the loop interferometer for polarization measurement and the time when the SNSPD outputs a signal.  
 
All the time-space relations are drawn to scale in Figure 3 of the main text. Therefore, further time-space relations can be inferred with the seven sets of space-like separation criteria. We use the parameters aforementioned and precisely measure them in the experiment. 
 
The size of each node (on an optical table) is less than 1 m. The beeline distances between the nodes have been precisely measured using a laser rangefinder with an uncertainty within 0.4 cm, much smaller than the node size. Therefore, we can place an upper bound of the distance uncertainty to be 1 m. The lengths of all the fibers have been measured beforehand. Taking into account the fiber fusion accuracy, the total fiber length uncertainty is within 0.1 m. The relevant beeline distance, fiber length between all nodes, and the relevant elapsed time with the beeline distances and their relative angles are shown in Table. \ref{tab:spacedistance} (a) and the fiber length between adjacent nodes are shown in Table. \ref{tab:fiberdistance} (b). The angles in Table. \ref{tab:spacedistance} (a) are calculated using the beeline distances.

The time elapse of the QRNGs to output a random bit is defined as the time interval between the start of the experiment trail and the output of an electrical signal from the FPGA. By using a signal generator to simulate the input of the QRNG, and measuring the output signal by an oscilloscope, we calibrated the time elapses to be less than 53 ns for 1-bit QRNG and 89 ns for 2-bits QRNG with a jitter within 2 ns.

Together, the detailed events are summarized in the Table. \ref{tab:spacetimedetails}. It is clear from Table. IV that all the seven set of space-like conditions are sufficiently fulfilled, ranging from $26\pm4$ ns to $1090\pm4$ ns, which ensures the truly independent quantum preparation and measurement, and the closing of the locality loopholes simultaneously.

  \begin{table}[!h]
 	\centering
 	\caption{The Spatial and fiber links between the stations in our experiment.}
 	\renewcommand{\arraystretch}{1.4}
 	\begin{minipage}{.5\linewidth}
 		\includegraphics[width=0.98\textwidth]{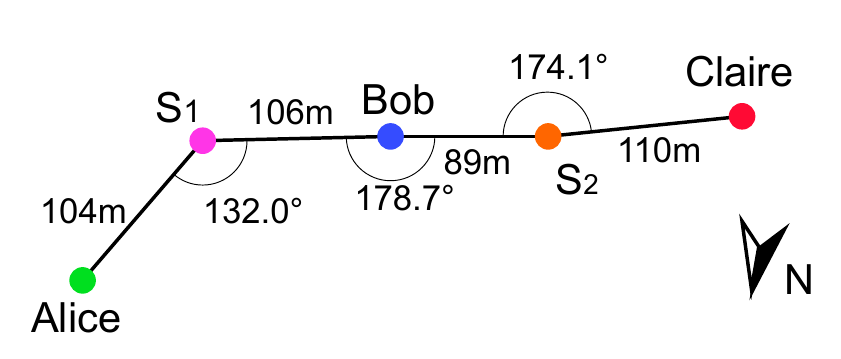}
    	\vspace{0.1cm} 
 		\\(a) The relative positions of the two sources and three measurement stations with the beeline distance. The upper bound of the uncertainty are 1 m.
 		\label{tab:spacedistance}
 \end{minipage}
 	\hfill
 		\begin{minipage}{.45\linewidth}
 		\begin{tabular}{|c|c|c|c|c|}
 			\hline
 			Fiber link & Alice-$\text{S}_{1}$ & $\text{S}_{1}$-Bob  & Bob-$\text{S}_{2}$  &  $\text{S}_{2}$-Claire \\ \hline
 			Length (m) & 112.63 & 123.84&108.75&125.48\\ \hline
 		\end{tabular}
 		\vspace{0.1cm} 
 		\\(b) The length of fiber links between adjacent nodes. The uncertainty are within 0.10 m.
 		\label{tab:fiberdistance}
 \end{minipage}
 \end{table}

\begin{table}[!t]
	\renewcommand{\arraystretch}{1.4}
	\centering
	\caption{The detailed space-like separation condition in our experiment. The subscripts, ``1" and ``2" are for $\text{S}_{1}$ and $\text{S}_{2}$; A, B, and C stand for Alice's, Bob's, and Claire's station; Aq, Bq and Cq represent the quantum random number generator at station for Alice, Bob, and Claire, respectively.}
	\begin{tabular}{|>{\centering}m{7cm}|>{\centering}m{1.5cm}|>{\centering}m{2cm}|>{\centering}m{1.3cm}|>{\centering}m{2cm}|c|}
		\hline
		Space-like separation constraints & \multicolumn{2}{c|}{Distances (m)} & \multicolumn{2}{c|}{Time (ns)} &  Results (ns) \\\hline
		
		\multirow{3}{*}{$\text{L}_{12}/c-\text{max}(t_{12}+\tau_{1},t_{12}+\tau_{2})>0$} & \multirow{3}{*}{$\text{L}_{12}$} & \multirow{3}{*}{195$\pm$1} & $t_{12}$ & 81.5$\pm$.7 & \multirow{3}{*}{408$\pm$4} \\ \cline{4-5}
		&   &   & $\tau_{1}$ & 160.2$\pm$0.5 & \\ \cline{4-5}
		&   &   & $\tau_{2}$ & 154.4$\pm$.5 & \\\hline
		
		\multirow{3}{*}{\shortstack{$\text{L}_{A1}/c-(t_{Aq1}+\tau_{Aq})>0$\vspace{0.3cm} \\$\text{L}_{A2}/c-(t_{Aq2}+\tau_{Aq})>0$}} & \multirow{3}{*}{\shortstack{$\text{L}_{A1}$\vspace{0.3cm} \\$\text{L}_{A2}$}} & \multirow{3}{*}{\shortstack{104$\pm$1\vspace{0.3cm} \\277$\pm$1}} & $t_{Aq1}$ & 225.7$\pm$0.7 & \multirow{3}{*}{\shortstack{32$\pm$4\vspace{0.3cm} \\690$\pm$4}} \\ \cline{4-5}
		&   &   & $t_{Aq2}$ & 144.2$\pm$0.7& \\ \cline{4-5}
		&   &   & $\tau_{Aq}$ & 89$\pm$2 & \\\hline
		
		\multirow{3}{*}{\shortstack{$\text{L}_{B1}/c-(t_{Bq1}+\tau_{Bq})>0$\vspace{0.3cm} \\$\text{L}_{B2}/c-(t_{Bq2}+\tau_{Bq})>0$}} & \multirow{3}{*}{\shortstack{$\text{L}_{B1}$\vspace{0.3cm} \\$\text{L}_{B2}$}} & \multirow{3}{*}{\shortstack{106$\pm$1\vspace{0.3cm} \\89$\pm$1}} & $t_{Bq1}$ & 274.2$\pm$0.7 & \multirow{3}{*}{\shortstack{26$\pm$4\vspace{0.3cm} \\51$\pm$4}} \\ \cline{4-5}
		&   &   & $t_{Bq2}$ & 192.8$\pm$0.7& \\ \cline{4-5}
		&   &   & $\tau_{Bq}$ & 53$\pm$2 & \\\hline
		
		\multirow{3}{*}{\shortstack{$\text{L}_{C1}/c-(t_{Cq1}+\tau_{Cq})>0$\vspace{0.3cm} \\$\text{L}_{C2}/c-(t_{Cq2}+\tau_{Cq})>0$}} & \multirow{3}{*}{\shortstack{$\text{L}_{C1}$\vspace{0.3cm} \\$\text{L}_{C2}$}} & \multirow{3}{*}{\shortstack{305$\pm$1\vspace{0.3cm} \\110$\pm$1}} & $t_{Cq1}$ & 327.1$\pm$0.7& \multirow{3}{*}{\shortstack{601$\pm$4\vspace{0.3cm} \\32$\pm$4}} \\ \cline{4-5}
		&   &   & $t_{Cq2}$ & 245.7 $\pm$0.7 & \\ \cline{4-5}
		&   &   & $\tau_{Cq}$ & 89$\pm$2 & \\\hline
		
		\multirow{3}{*}{$\text{L}_{AC}/c-\text{max}(t_{AqCq}+\tau_{Aq},t_{AqCq}+\tau_{Cq})>0$} & \multirow{3}{*}{$\text{L}_{AC}$} & \multirow{3}{*}{384$\pm$1} & $t_{AqCq}$ & 101.5$\pm$0.7 & \multirow{3}{*}{1090$\pm$4} \\ \cline{4-5}
		&   &   & $\tau_{Aq}$ & 89$\pm$2 & \\ \cline{4-5}
		&   &   & $\tau_{Cq}$ & 89$\pm$2 & \\\hline
		
		\multirow{3}{*}{$\text{L}_{BA}/c-\text{max}(t_{BqAq}+\tau_{Bq},t_{BqAq}+\tau_{Aq})>0$} & \multirow{3}{*}{$\text{L}_{BA}$} & \multirow{3}{*}{192$\pm$1} & $t_{BqAq}$ & 48.6$\pm$0.7 & \multirow{3}{*}{502$\pm$4} \\ \cline{4-5}
		&   &   & $\tau_{Bq}$ & 53$\pm$2 & \\ \cline{4-5}
		&   &   & $\tau_{Aq}$ & 89$\pm$2 & \\\hline
		
		\multirow{3}{*}{$\text{L}_{BC}/c-\text{max}(t_{BqCq}+\tau_{Bq},t_{BqCq}+\tau_{Bq})>0$} & \multirow{3}{*}{$\text{L}_{BC}$} & \multirow{3}{*}{199$\pm$1} & $t_{BqCq}$ & 52.9$\pm$0.7 & \multirow{3}{*}{521$\pm$4} \\ \cline{4-5}
		&   &   & $\tau_{Bq}$ & 53$\pm$2 & \\ \cline{4-5}
		&   &   & $\tau_{Cq}$ & 89$\pm$2 & \\\hline

		\multirow{4}{*}{\shortstack{$\text{L}_{AB}/c- (t_{AqB}+\tau_B)>0$\vspace{0.4cm} \\$\text{L}_{AC}/c- (t_{AqC}+\tau_C)>0$}} & \multirow{4}{*}{\shortstack{$\text{L}_{AB}$\vspace{0.4cm} \\$\text{L}_{AC}$}} & \multirow{4}{*}{\shortstack{192$\pm$1\vspace{0.4cm} \\384$\pm$1}} & $t_{AqB}$ & 553.6$\pm$0.7& \multirow{4}{*}{\shortstack{42$\pm$4\vspace{0.4cm} \\604$\pm$4}} \\ \cline{4-5}
		&   &   & $\tau_{B}$ & 44.9$\pm$0.5& \\ \cline{4-5}
		&   &   & $t_{AqC}$ & 637.3$\pm$0.7 & \\ \cline{4-5}
		&   &   & $\tau_{C}$ & 38.4$\pm$0.5 & \\\hline

		\multirow{4}{*}{\shortstack{$\text{L}_{BA}/c- (t_{BqA}+\tau_A)>0$\vspace{0.4cm} \\$\text{L}_{BC}/c- (t_{BqC}+\tau_C)>0$}} & \multirow{4}{*}{\shortstack{$\text{L}_{BA}$\vspace{0.4cm} \\$\text{L}_{BC}$}} & \multirow{4}{*}{\shortstack{192$\pm$1\vspace{0.4cm} \\199$\pm$1}} & $t_{BqA}$ & 449.0$\pm$0.7& \multirow{4}{*}{\shortstack{146$\pm$4\vspace{0.4cm} \\36$\pm$4}} \\ \cline{4-5}
		&   &   & $\tau_{A}$ & 44.6$\pm$0.5 & \\ \cline{4-5}
		&   &   & $t_{BqC}$ & 588.7$\pm$0.7 & \\ \cline{4-5}
		&   &   & $\tau_{C}$ & 38.4$\pm$0.5 & \\\hline
		
		\multirow{4}{*}{\shortstack{$\text{L}_{CA}/c- (t_{CqA}+\tau_A)>0$\vspace{0.4cm} \\$\text{L}_{CB}/c- (t_{CqB}+\tau_B)>0$}} & \multirow{4}{*}{\shortstack{$\text{L}_{CA}$\vspace{0.4cm} \\$\text{L}_{CB}$}} & \multirow{4}{*}{\shortstack{384$\pm$1\vspace{0.4cm} \\199 $\pm$1}} & $t_{CqA}$ & 396.1$\pm$0.7& \multirow{4}{*}{\shortstack{839$\pm$4\vspace{0.4cm} \\166$\pm$4}} \\ \cline{4-5}
		&   &   & $\tau_{A}$ & 44.6$\pm$ 0.5 & \\ \cline{4-5}
		&   &   & $t_{CqB}$ & 452.1$\pm$0.7 & \\ \cline{4-5}
		&   &   & $\tau_{B}$ & 44.9$\pm$0.5 & \\\hline
		
	\end{tabular}
	\label{tab:spacetimedetails}
\end{table}

\end{document}